\documentclass[prl,aps,twocolumn]{revtex4-1}

\usepackage{graphicx}
\usepackage{bm}
\usepackage{amsfonts}
\usepackage{color}
\usepackage[breaklinks]{hyperref}
\usepackage{epstopdf, epsfig}

\newcommand{\de}{\mathrm{d}}
\newcommand{\mubar}{\mathcal{F}}
\newcommand{\mustar}{\mathcal{F}^\star}
\newcommand{\Rey}{\mathrm{Re}}
\newcommand{\Lf}{\ell}
\newcommand{\Xg}{\bar{\bm X}}
\newcommand{\dd}[2]{\frac{\de{#1}}{\de{#2}}}

\newcommand{\ocaaddress}{Universit\'e C\^ote d'Azur, CNRS, OCA,
  Laboratoire J.-L.\ Lagrange, Nice, France}

\begin{document}

\title{Stretching and Buckling of Small Elastic Fibers in Turbulence}

\author{Sof\'{\i}a Allende} \affiliation{\ocaaddress}
\author{Christophe Henry} \affiliation{\ocaaddress}
\author{J\'er\'emie Bec} \affiliation{\ocaaddress}

\begin{abstract}
  Small flexible fibers in a turbulent flow are found to be most of
  the time as straight as stiff rods. This is due to the cooperative
  action of flexural rigidity and fluid stretching.  However, fibers
  might bend and buckle when they tumble and experience a
  strong-enough local compressive shear. Such events are similar to an
  activation process, where the role of temperature is played by the
  inverse of Young's modulus.  Numerical simulations show that buckling
  occurs very intermittently in time. This results from unexpected
  long-range Lagrangian correlations of the turbulent shear.
\end{abstract}

\maketitle

Elongated colloidal particles are essentially subject to three
dynamical forces: Bending elasticity, thermal fluctuations, and
viscous drag with the suspending flow.  An important and well-studied
case is that of infinitely flexible polymers for which only two
effects compete: Coiling promoted by thermal noise, and stretching
induced by fluid shear.  Relaxation to equilibrium is then fast enough
to give grounds for adiabatic macroscopic models, such as elastic
dumbbells~\cite{deGennes1979scaling,watanabe2010coil}, often used to
investigate the rheology of polymer
suspensions~\cite{pereira2017active}. Much less is known when the
thermal fluctuations are negligible but bending elasticity becomes
important.  This asymptotics is relevant to describe macroscopic
particles, such as cellulose fibers in papermaking
industry~\cite{lundell2011fluid}, or diatom phytoplankton
colonies~\cite{ardekani2017sedimentation} that significantly
participate to the CO$_2$ oceanic pump~\cite{smetacek1999diatoms}. In
principle, without molecular diffusion there is no coiling.
Furthermore, bending elasticity and flow strain act concomitantly to
stretch the fiber, suggesting an unsophisticated stiff rod dynamics.
However, most natural or industrial flows are turbulent. They thus
display violent and intermittent fluctuations of velocity gradients,
susceptible of destabilizing a straight configuration and leading to
the buckling of the fiber~\cite{becker2001instability}.

We are interested in elongated, deformable, macroscopic particles
passively transported by a turbulent flow.  We aim at quantifying two
aspects: First, the extent to which their dynamics can be approximated
as that of rigid rods and, second, the statistics of buckling.  For
that purpose, we focus here on the simplest model, the \emph{local}
slender-body theory (see, e.g., \cite{lindner2015elastic}), which
describes flexible fibers with cross-section $a$ and length $\Lf$, as
\begin{eqnarray}
  \partial_t \bm X &=& \bm u(\bm X,t)+
  \frac{c}{8\pi\,\rho_{\rm f}\,\nu}\,\mathbb{D}\,
  \left[\partial_s(T\,\partial_s \bm X) - E\,\partial_s^4\bm X\right],
                       \nonumber\\
  && |\partial_s\bm X|^2 = 1, \ \mbox{with } \mathbb{D} =
  \mathbb{I} + \partial_s\bm X\,\partial_s \bm X^{\mathsf{T}},
  \label{eq:SBT}
\end{eqnarray}
in the asymptotics $c = -[1+2\,\log (a/\Lf)] \gg 1$.  Here,
$\bm X(s,t)$ is the spatial position of the point indexed by the arc
length coordinate $s\in[-\Lf/2,\Lf/2]$; $\bm u$ is the velocity field
of the fluid, $\nu$ its kinematic viscosity and $\rho_{\rm f}$ its
mass density; $E$ denotes the fiber's Young modulus. The tension
$T(s,t)$, which satisfies $T|_{\pm \Lf/2} = 0$, is the Lagrange
multiplier associated to the fiber's inextensibility constraint.
Equation~(\ref{eq:SBT}) is supplemented by the free-end boundary
conditions $\partial_s^2\bm X|_{\pm \Lf/2} = 0$ and
$\partial_s^3\bm X|_{\pm \Lf/2} =0$.

The considered fibers are much smaller than the smallest active scale
of the fluid velocity $\bm u$. In turbulence, this means $\Lf\ll\eta$,
where $\eta = \nu^{3/4}/\varepsilon^{1/4}$ is the Kolmogorov
dissipative scale,
$\varepsilon \!= \!\nu \langle \|\nabla \bm u\|^2 \rangle$ being the
turbulent rate of kinetic energy dissipation.  In this limit, the
particle motion is to leading order that of a tracer and
${\de\Xg}/\de t = \bm u( \Xg ,t)$, where $\Xg(t)$ denotes its center
of gravity. The deformation of the fiber solely depends on the local
velocity gradient, so that
$\bm u(\bm X,t) \approx \bm u( \Xg ,t)+\mathbb{A}(t)\,(\bm
X\!-\!\Xg)$, where $\mathbb{A}_{ij}(t) = \partial_j u_i(\Xg ,t)$. The
dynamics is then fully described by two parameters: The fluid flow
Reynolds number $\Rey$, prescribed very large, and the non-dimensional
fiber flexibility
\begin{equation}
  \mubar = \frac{8\pi\,\rho_{\rm f}\,\nu\,\Lf^4}{c\,E\,\tau_\eta},
\end{equation}
where $\tau_\eta = \sqrt{\nu/\varepsilon}$ is the Kolmogorov
dissipative time and quantifies typical values of the turbulent
strain rate. The parameter $\mubar$ can be understood as the ratio
between the timescale of the fiber's elastic stiffness to that of the
turbulent velocity gradients. At small $\mubar$, the fiber is very
rigid and always stretched. On the contrary, for large $\mubar$ it is
very flexible and might bend.

In the fully-stretched configuration, the tangent vector is constant
along the fiber, i.e.\ $\partial_s\bm X = \mathbf{p}(t)$, and follows
Jeffery's equation for straight ellipsoidal
rods~\cite{tornberg2004simulating}
\begin{equation}
  \dd{\mathbf{p}}{t} = \mathbb{A}\,\mathbf{p} -
  (\mathbf{p}^{\mathsf{T}}\mathbb{A}\,\mathbf{p})\,\mathbf{p}.
  \label{eq:Jeffery}
\end{equation}
This specific solution to (\ref{eq:SBT}), which is independent of $s$,
is stable when the fiber is sufficiently rigid.  However, it becomes
unstable when increasing flexibility, or equivalently for larger fluid
strain rates.  As shown and observed experimentally in two-dimensional
velocity fields, such as linear
shear~\cite{becker2001instability,harasim2013direct,liu2018tumbling}
or extensional
flows~\cite{young2007stretch,wandersman2010buckled,kantsler2012fluctuations},
this instability is responsible for a buckling of the fiber.  This
occurs when the elongated fiber
tumbles~\cite{schroeder2005characteristic,plan2016tumbling} and
experiences a strong-enough compression along its direction.  This
compression is measured by projecting the velocity gradient along the
rod directions, i.e.\ by the shear rate
$\dot{\gamma} = \mathbf{p}^{\mathsf{T}}\mathbb{A}\,\mathbf{p}$.  In
turbulence, buckling thus occurs when the instantaneous value of
$\dot{\gamma}$ becomes large with a negative value (compression).

To substantiate this picture we have performed direct numerical
simulations.  The flow is obtained by integrating the incompressible
Navier--Stokes equations using the \emph{LaTu} spectral solver with
$4096^3$ collocation points and with a force maintaining constant the
kinetic energy content of the two first Fourier shells (see, e.g.,
\cite{homann2007impact}).  Once a statistically stationary state is
reached with a Taylor micro-scale Reynolds number
$\Rey_\lambda\approx730$, the flow is seeded with several thousands of
tracers, along which the full velocity gradient tensor is stored with
a period $\approx\tau_\eta/4$.  The local slender-body equation for
fibers~(\ref{eq:SBT}) is then integrated a posteriori along these
tracer trajectories. We use the semi-implicit, finite-difference
scheme introduced in~\cite{tornberg2004simulating}, with the
inextensibility constraint enforced by a penalization procedure.
$N = 201$ grid points are used along the fibers arc-length coordinate,
with a time step $5\times10^{-4}\,\tau_\eta$. We use linear
interpolation in time to access the velocity gradient at a higher
frequency than the output from the fluid simulation.

\begin{figure}[b]
  \begin{center}
    \includegraphics[width=.97\columnwidth]{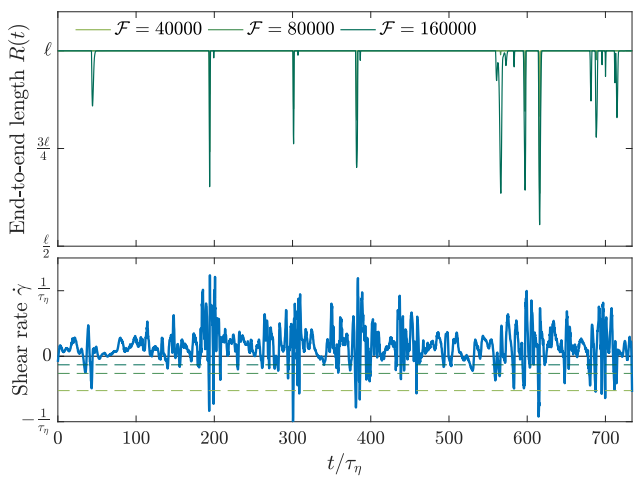}
  \end{center}
  \vspace{-18pt}
  \caption{\label{fig:traj} Top panel: Time evolution of the
    end-to-end length $R(t)$ for a specific turbulent tracer
    trajectory and three different values of the non-dimensional
    flexibility $\mubar$, as labeled. Bottom panel: Evolution of the
    instantaneous shear rate $\dot{\gamma}$ along the same
    trajectory. The solid line corresponds to $\dot{\gamma}=0$ and the
    dashed lines to $\tau_\eta\,\dot{\gamma}=-0.13$, $-0.26$, and
    $-0.52$. Note that time is rescaled by the Kolmogorov timescale
    $\tau_\eta$. In these units, the large-eddy turnover time is
    $\tau_L\approx 190\,\tau_\eta$.}
\end{figure}
Numerics confirm that fibers much smaller than the Kolmogorov scale
are almost always stretched. This can be measured from the end-to-end
length $R(t) = |\bm X(\ell/2,t) - \bm X(-\ell/2,t)|$. When $R=\ell$,
the fiber is straight. Buckling occurs when $R<\ell$. The upper panel
of Fig.~\ref{fig:traj} shows the time evolution of the end-to-end
length along a single trajectory for various non-dimensional
flexibilities $\mubar$. Clearly, bending is sparse and
intermittent. Buckling events are separated by long periods during
which $R\equiv\ell$, up to numerical precision. For instance, one
observes $|1-R(t)/\ell| <10^{-13}$ in the time interval
$100<t/\tau_\eta<180$. In the lower panel of Fig.~\ref{fig:traj}, we
have shown the time evolution of the shear rate $\dot{\gamma}$ along
the same Lagrangian trajectory. As expected, buckling events are
associated to strong negative fluctuations of $\dot{\gamma}$. Note
that, because $\mathbf{p}$ is preferentially aligned with the fluid
stretching~\cite{ni2014alignment}, the shear rate has a positive mean
$\langle \dot{\gamma} \rangle \approx 0.11/\tau_\eta$. Its standard
deviation is $\approx 0.2/\tau_\eta$.

\begin{figure}[b]
  \begin{center}
    \includegraphics[width=\columnwidth]{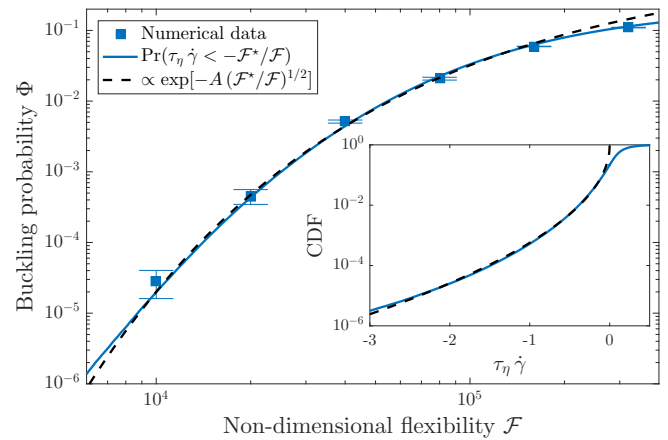}
  \end{center}
  \vspace{-18pt}
  \caption{\label{fig:prob_buckl} Probability of buckling $\Phi$ as a
    function of $\mubar$. The squares come from numerical simulations
    and were calculated as the fraction of time during which
    $R(t)/\ell<0.999$. Error bars are obtained from the standard
    deviation of the duration of individual events. The bold line is
    the probability that the shear rate $\dot{\gamma}$ is less than
    $-\mustar/(\tau_\eta\,\mubar)$ with $\mustar = 2.1\times
    10^4$. The corresponding cumulative distribution function (CDF) is
    shown in the inset, together with a fit (dashed line) of the form
    $\propto\exp[-A\,(\tau_\eta\,|\dot{\gamma}|)^{1/2}]$, where
    $A=1.69$. The same approximation is used in the main panel to fit
    $\Phi$ (dashed line).}
\end{figure}
To get more quantitative insights, we define buckling events as times
when $R(t)/\ell$ is below a prescribed threshold (we have used
$0.999$). Figure~\ref{fig:prob_buckl} shows the probability of
buckling as a function of the flexibility $\mubar$. This quantity,
denoted $\Phi$, is defined as the fraction of time spent by the
end-to-end length below this threshold.  One finds that, conversely to
simple steady shear flows (see, e.g., \cite{lindner2015elastic}), in
turbulence there is no critical value of the flexibility above which
buckling occurs.  Fibers bending is similar to an activated process
with $\Phi \propto \exp(-C/\mubar^\alpha)$ and where the flexibility
$\mubar$ plays a role resembling that of temperature in chemical
reactions. Indeed, the fiber will buckle when its instantaneous
flexibility
$\mubar_{\rm loc}(t) = \tau_\eta\,|\dot{\gamma}(t)|\,\mubar$ is larger
than a critical value $\mustar$, with $\dot{\gamma}(t)<0$. This leads
to
\begin{equation}
  \Phi = \mathrm{Pr}\,\left(\tau_\eta\,\dot{\gamma}< -\mustar/\mubar
  \right).
\end{equation}
As can be seen in Fig.~\ref{fig:prob_buckl}, the cumulative
probability of the shear rate can indeed be used to reproduce the
numerical measurements of $\Phi$ by choosing
$\mustar = 2.1\times 10^4$. This value, which just corresponds to a
fit, is much larger than those observed in time-independent shear
flows~\cite{liu2018tumbling} where buckling occurs for
$\mubar\gtrsim 300$.  A first reason comes from using the Kolmogorov
dissipative timescale when defining $\mubar$. This is a natural but
arbitrary choice in turbulence. However $\tau_\eta$ is significantly
smaller than typical values of $\dot{\gamma}^{-1}$, so that effective
flexibilities could be smaller than $\mubar$.  This is similar to
choosing $\tau_\eta$ rather than the Lyapunov exponent to define the
Weissenberg number for the coil-stretch transition of dumbbells in
turbulent flow~\cite{boffetta2003two}.  Another explanation for a
large $\mustar$ could be the intricate relation in turbulence between
the amplitude of velocity gradients and their dynamical timescales,
which implies in principle that the stronger is $\dot{\gamma}$, the
shorter is the lifetime of the associated velocity gradient.

Fibers with a small flexibility buckle only when the instantaneous
shear rate is sufficiently violent. Moreover it is known that at large
Reynolds numbers~\cite{kailasnath1992probability}, the probability
distribution of velocity gradients has stretched-exponential tails
with exponent $\approx 1/2$.  This behavior is also present in the
cumulative probability of $\dot{\gamma}$, as seen in the inset of
Fig.~\ref{fig:prob_buckl}.  This leads to predict that
\begin{equation}
  \Phi \propto \mathrm{e}^{
    -A\,\left({\mustar}/{\mubar}\right)^{1/2}}
  \mbox{ for }\mubar \ll \mustar.
\end{equation}
This asymptotic behavior is shown as a dashed line in the main panel
of Fig.~\ref{fig:prob_buckl}. It gives a rather good fit of the data,
up to $\mubar\approx1.6\times 10^5$. At larger values, this
activation-like asymptotics and relation to the tail of the
distribution is no more valid.  At very small values (or equivalently
large negative $\dot{\gamma}$'s), one observes tiny deviations from
the stretched exponential, certainly resulting from numerical errors
over-predicting extreme gradients~\cite{buaria2017resolution}.

The relevance to buckling of an instantaneous flexibility larger than
$\mustar$ can be seen in Fig.~\ref{fig:traj}. The dashed lines on the
bottom panel are the critical values
$\tau_\eta\,\dot{\gamma} =-\mustar/\mubar$ associated to the three
flexibilities of the top panel.  We indeed observe that buckling
occurs when the instantaneous shear rate underpasses these values. In
some cases (e.g.\ for times between $400$ and $500\,\tau_\eta$), it
seems that the fiber is straight, even if $\dot{\gamma}$ is below the
threshold. Still, buckling occurs but with an amplitude so small that
it cannot be detected from the top panel.  This threshold therefore
provides information on the occurence of buckling, but not on the
strength of the associated bending.

\begin{figure}[t]
  \begin{center}
    \includegraphics[width=\columnwidth]{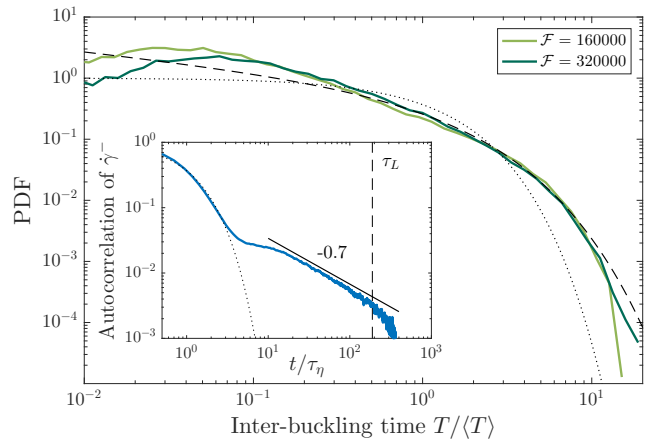}
  \end{center}
  \vspace{-18pt}
  \caption{\label{fig:waiting_time} Probability density functions
    (solid lines) of the time $T$ between successive buckling events,
    normalized to its average $\langle T\rangle \approx 52\,\tau_\eta$
    for $\mubar=1.6\times10^5$, and
    $\langle T\rangle \approx 36\,\tau_\eta$ for
    $\mubar=3.2\times10^5$. The dotted line represents the exponential
    distribution. The dashed line is a Weibull distribution
    (\ref{eq:weibull}) with shape parameter $\beta=0.7$ and scale
    parameter $\lambda = 1$. Inset: Autocorrelation
    $\rho(t) = \mathrm{cov} \left(\dot{\gamma}^-(t),
      \dot{\gamma}^-(0)\right) / \mathrm{Var}(\dot{\gamma}^-)$ of the
    negative part of the shear rate. The dotted line stands for
    $\exp(-t/\tau_\eta)$; The vertical dashed line indicates the
    large-eddy turnover time $t=\tau_L$; The solid line shows a slope
    $-0.7$.}
\end{figure}

Another qualitative assessment that can be drawn from
Fig.~\ref{fig:traj} is that large excursions of $\dot{\gamma}$ are not
isolated events but form clumps.  This is a manifestation of the
Lagrangian intermittency of velocity gradients.  Tracers might indeed
be trapped for long times in excited regions of the flow, leading to
fluctuations correlated over much longer times than $\tau_\eta$. This
can be quantified from the autocorrelation $\rho(t)$ of the negative
part $\dot{\gamma}^- = \max(-\dot{\gamma},0)$ of the shear rate, which
is represented in the inset of Fig.~\ref{fig:waiting_time}. The
corresponding integral correlation time is
$\int\rho(t)\,\mathrm{d}t \approx 2.8\,\tau_\eta$.  This can be
explained by the abrupt decrease of the autocorrelation at times of
the order of the Kolmogorov timescale.  This behavior is essentially a
kinematic effect due to fast rotations.  Remember that $\dot{\gamma}$
is obtained by projecting the velocity shear on the direction
$\mathbf{p}$ of a rigid rod. This direction rotates with an angular
speed given by the vorticity $\omega = |\nabla\times\bm u|$, so that
$\dot{\gamma}$ can alternate from expansion to compression, on
timescales of the order of $\omega^{-1}\sim\tau_\eta$.  Surprisingly,
at longer times $t \gtrsim 4\,\tau_\eta$, the autocorrelation of
$\dot{\gamma}^-$ changes regime and decreases much slower than an
exponential.  This contradicts the classical phenomenological vision
that velocity gradients are purely a small-scale quantity with
correlations spanning only the dissipative scales. For more than a
decade in $t$ within the inertial range, we indeed find a power-law
behavior $\rho(t)\propto t^{-\beta}$, with
$\beta \approx 0.7 \pm 0.1$. To our knowledge, this is the first time
such a long-range behavior is observed for turbulent Lagrangian
correlations.

These intricate correlations have important consequences on the
incidence of buckling.  Memory effects are present, as the fiber is
likely to bend several times when in a clump of violent,
high-frequency fluctuations of $\dot{\gamma}$.  Consequently, the
probability distribution $p(T)$ of the time $T$ between successive
buckling events is not an exponential. This is clear from the main
panel of Fig.~\ref{fig:waiting_time}, where this distribution is shown
for $\mubar=1.6\times10^5$ and $3.2\times10^5$. One observes clear
deviations from the exponential distribution (dotted line).  They
relate to the two regimes discussed above for the time correlations of
$\dot{\gamma}^-$.  First, the distribution of inter-buckling times is
maximal for $T$ of the order of $\tau_\eta$.  This corresponds to
rapid oscillations of the sign of $\dot{\gamma}$.  The fiber
experiences several tumblings in an almost-constant velocity gradient
and is alternatively compressed and pulled out by the flow due to fast
rotations. This leads to a rapid succession of bucklings and
stretchings.  Second, strong deviations to the exponential
distribution also occur for inter-buckling times $T$ in the inertial
range.  As seen in Fig.~\ref{fig:waiting_time}, the distribution of
inter-buckling times in the intermediate range
$0.5 \lesssim T/\langle T\rangle \lesssim 5$ is well approximated by a
Weibull distribution with shape $\beta$ and scale parameter $\lambda$:
\begin{equation}
  p(T) \approx \frac{\beta\,T^{\beta-1}}{\lambda^\beta}\,
  \mathrm{e}^{-\left({T}/{\lambda}\right)^\beta}.
  \label{eq:weibull}
\end{equation}
This decade exactly matches the time lags for which $\dot{\gamma}$
displays long-range correlations, that is $\rho(t)\sim t^{-\beta}$.
The return statistics of processes with power-law correlations is
indeed expected to be well approximated by a Weibull
distribution~\cite{eichner2007statistics}.  Longer times correspond to
$t\gtrsim\tau_L$, for which $p(T)$ is expected to ultimately approach
an exponential tail.

\begin{figure}[h]
  \begin{center}
    \includegraphics[width=\columnwidth]{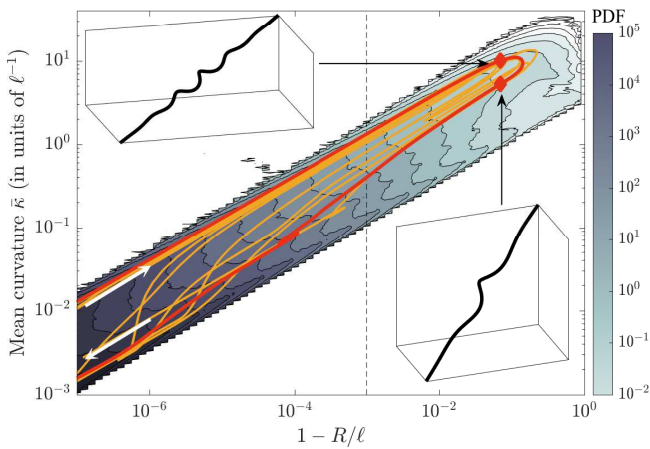}
  \end{center}
  \vspace{-18pt}
  \caption{\label{fig:buckling_stat} Contour levels of the joint
    distribution of the end-to-end length $R$ and of the mean
    curvature $\bar{\kappa}$ for $\mubar = 1.6 \times 10^5$. The vertical
    dashed line shows the threshold $R/\ell = 0.999$.  Two excursions
    of the same trajectory as Fig.~\ref{fig:traj} are shown for
    $43<t/\tau_\eta<77$ (red) and $680<t/\tau_\eta<730$
    (orange). Also, two instantaneous configurations of the fiber are
    represented for $t=43\,\tau_\eta$ (top left) and $t=45\,\tau_\eta$
    (bottom right).}
\end{figure}
To characterize further buckling events and in particular their
geometry, we show in Fig.~\ref{fig:buckling_stat} the joint
probability density of the end-to-end length $R$ and of the fiber's
mean curvature
$\bar{\kappa} = (1/\ell)\int |\partial_s^2\bm X|\,\mathrm{d}s$. The
distribution is supported in a thin strip aligned with
$\bar{\kappa} \propto (1-R/\ell)^{1/2}$. Bucklings correspond to loops
in this plane. Trajectories typically start such excursions with a
larger curvature (upper part of the strip) than the one they have when
relaxing back to a straight configuration (lower part). The red curve
corresponds to the first buckling event of the trajectory shown in
Fig.~\ref{fig:traj}. The curvature increases concomitantly to a
decrease of the end-to-end length. Right before reaching a maximal
bending, the fiber displays several coils (top-left inset).  This
configuration depends on the most unstable mode excited with the
current value of the instantaneous flexibility $\mubar_{\rm loc}$. It
is indeed known that buckling fibers in steady shear flows can
experience several bifurcations, depending on their
elasticity~\cite{becker2001instability}. Once the fiber has again
aligned with $\dot{\gamma}>0$, that is a couple of $\tau_\eta$'s
later, these coils unfold (bottom-right inset), the curvature
decreases, and the fiber relaxes back to a fully stretched
configuration. This specific event has been chosen for its simplicity
and representativity.  In the case of high-frequency buckling
discussed earlier, such as shown in orange, the fiber experiences
multiple buckling. The time is now $t\approx 700\,\tau_\eta$, for
which we observe in Fig.~\ref{fig:traj}, several oscillations of
$\dot{\gamma}$. The fiber is alternatively compressed and stretched,
leading here to six successive bendings, separated by only a few
$\tau_\eta$'s.

To conclude, recall that we have focused on passively transported
fibers.  In several applications, they actually have an important
feedback on the flow and might even reduce turbulent drag. We found
here that the dynamics of flexible fibers strongly depends on the
shear strength: In calm regions, they just behave as stiff rods.  In
violent, intermittent regions, they can buckle, providing an effective
transfer of kinetic energy toward bending elasticity. Such nonuniform,
shear-dependent effects likely lead to intricate flow modifications,
where the presence of small fibers affects not only the amplitude of
turbulent fluctuations, but also their very nature.  Such complex
non-Newtonian effects undoubtedly lead to novel mechanisms of
turbulence modulation.

This work has been supported by EDF R\&D (projects PTHL of MFEE and
VERONA of LNHE) and by the French government,
through the Investments for the Future project UCA$^{\mathrm{JEDI}}$
ANR-15-IDEX-01 managed by the Agence Nationale de la Recherche.

\end{document}